\documentclass[pre,aps,twocolumn, superscriptaddress]{revtex4}
\usepackage{amsmath}
\usepackage{epsfig}
\usepackage{amssymb}
\usepackage{color}

\begin{document}

\title{Noise-induced drift in two-dimensional anisotropic systems}

\author{Oded Farago} 
\affiliation{Department of Biomedical
Engineering, Ben-Gurion University of the Negev, Be'er Sheva 85105,
Israel} 
\affiliation{Ilse Katz Institute for Nanoscale Science and
Technology, Ben-Gurion University of the Negev, Be'er Sheva 85105,
Israel}

\begin{abstract}

  We study the isothermal Brownian dynamics of a particle in a system
  with spatially varying diffusivity. Due to the heterogeneity of the
  system, the particle's mean displacement does not vanish even if it
  does not experience any physical force. This phenomenon has been
  termed ``noise-induced drift'', and has been extensively studied for
  one-dimensional systems. Here, we examine the noise-induced drift in
  a two-dimensional anisotropic system, characterized by a symmetric
  diffusion tensor with unequal diagonal elements.  A general
  expression for the mean displacement vector is derived and presented
  as a sum of two vectors, depicting two distinct drifting
  effects. The first vector describes the tendency of the particle to
  drift toward the high diffusvity side in each orthogonal principal
  diffusion direction. This is a generalization of the well-known
  expression for the noise-induced drift in one-dimensional
  systems. The second vector represents a novel drifting effect, not
  found in one-dimensional system, originating from the spatial
  rotation in the directions of the principal axes. The validity of
  the derived expressions is verified by using Langevin dynamics
  simulations. As a specific example, we consider the relative
  diffusion of two transmembrane proteins, and demonstrate that the
  average distance between them increases at a surprisingly fast rate
  of several tens of micrometers per second.

\end{abstract} 

\maketitle

\section {Introduction}
\label{sec:intro}

Recent advances in single particle tracking methods and force
measurement techniques have led to renewed interest in the problem of
isothermal Brownian motion in inhomogeneous systems \cite{volpe1}. A
prominent example is diffusion of a colloidal particle near a surface,
in which case due to hydrodynamic interactions, the diffusion
coefficients parallel and perpendicular to the boundary are (i)
different of each other, and (ii) increase with the particle-wall
distance \cite{brenner1,brenner2}. Another example is diffusion in
liquid crystalline systems where the orientation of the media is
heterogeneous \cite{tauber,pumpa}. Closely-related is the problem of
Brownian motion of non-spherical particles, e.g., ellipsoids whose
diffusion coefficients along the long and short axes are different
\cite{perrin,han}.

A common feature in the above examples of Brownian motion with
state-dependent diffusion is the {\em anisotropic}\/ nature of the
dynamics, i.e., the fact it is direction-dependent. Such problems
arise only in two and higher dimensions. Current theoretical
understanding of the topic of heterogeneous diffusion, however, is
based on studies of one-dimensional systems where a particle moves in
a medium with coordinate-dependent diffusion coefficient $D(x)$ (see
recent review \cite{volpe2}, and many references therein). The
dynamics of a freely diffusing particle (i.e., experiencing no
potential energy gradient) in a one-dimensional isothermal system, can
be described by Langevin's equation \cite{langevin}
\begin{eqnarray}
  m\frac{dv}{dt}=-\alpha(x)v+\beta\left(x\left( t\right)\right),
  \label{eq:langevineq}
\end{eqnarray}
where $m$ and $v=dx/dt$ denote, respectively, the mass and the
velocity of the particle. This is Newton's second law of motion where
the contact with the heat bath is realized via the action of two
effective forces: A friction force, $-\alpha v$, proportional to the
velocity with a coordinate-dependent friction coefficient
$\alpha(x)>0$, and a stochastic force $\beta(t)$ that can be modeled
as a multiplicative Gaussian noise with zero mean $\langle
\beta(t)\rangle=0$ and $\delta$-function auto-correlation
$\langle\beta(t)\beta(t^{\prime})\rangle=2\alpha(x(t))k_BT\delta(t-t^{\prime})$,
where $k_B$ is Boltzmann's constant and $T$ is the temperature of the
system. These statistical properties ensure that the
fluctuation-dissipation theorem is obeyed, which is necessary for
achieving correct Fickian dynamics \cite{kubo}. The state-dependent
diffusion coefficients $D(x)$ and $\alpha(x)$ are related to each
other via Einstein's relation \cite{gjf1}
\begin{eqnarray}
 \alpha(x)=k_BT/D(x).
  \label{eq:einstein}
\end{eqnarray}

The trajectory of the particle can be calculated by numerically
integrating Langevin's equation in time. From an ensemble of
stochastic trajectories, the probability distribution function (PDF),
$P(x,t)$, of finding the particle at coordinate $x$ at time $t$ can be
determined for a given initial distribution $P(x,0)$. In heterogeneous
systems with spatially-varying friction coefficient $\alpha(x)$,
Langevin's equation must be supplemented with a convention for
choosing the value $\alpha(x)$ at each integration time step $dt$. The
ambiguity about the appropriate convention rule is known in the
literature as the It\^{o}-Stratonovich dilemma \cite{mannella}. In the
``overdamped'' limit of Langevin's equation, which is when the
inertial term on the l.h.s.~of Eq.~(\ref{eq:langevineq}) is set
identically to zero, different conventions lead to different dynamics
and, consequently, different PDFs \cite{lau}. Keeping the inertial
term in Eq.~(\ref{eq:langevineq}), on the other hand, ensures that the
correct PDF is obtained when the integration time step $dt\rightarrow
0$, ragradless of the interpretation of the stochastic calculus
\cite{gjf2}. This remarkable difference between the underdamped and
overdamped Langevin equations is directly related to the most striking
feature of heterogeneous Brownian dynamics - the noise-induced drift.

The term ``noise-induced drift'' refers to the phenomenon that a
particle, freely diffusing in a medium with a coordinate-dependent
friction coefficient $\alpha(x)$, tends to drift toward the less
viscous side of the system \cite{volpe1}. By ``freely'' we mean in the
absence of external forces, concentration or temperature
gradients. The presence of a diffusion coefficient gradient allows for
a drift in the position of an individual particle {\em without a net
  particle current}\/ - a rather counterintuitive equilibrium
effect. The drift originates from the fact that when the particle
moves in the less viscous direction, it suffers less dissipation and
therefore travels longer distances \cite{gjf2}. The drifting effect is
countered by the tendency of the particle to diffuse more slowly and
get trapped for longer durations at the more viscous part of the
system. In closed system, the consequence of the opposite ``drifting"
and ``trapping" effects is the proper uniform equilibrium distribution
\cite{lancon}. We notice, however, that the drift is an inertial
effect taking place at short time scales $t\lesssim \tau=m/\alpha$
during which Langevin's dynamics is ballistic in nature. This effect
is missing in the overdamped version of Eq.~(\ref{eq:langevineq})
which, therefore, must be written with an additional term to correctly
account for the drifting effect. The magnitude of this, so called,
``spurious drift'' term depends on the chosen interpretation
\cite{lau}. The term spurious drift is, of course,
  misleading as the drift is a very real physical phenomenon.

In dimensions higher than one, the system is characterized by a
symmetric diffusion matrix that can be diagonalized along the
principal diffusion directions. It is expected to observe
noise-induced drift toward lower viscosity along each of these
principal directions, since the multidimensional dynamics decouples
into independent one-dimensions problems. This, however, is true only
if the directions of the principal axes are fixed. If, on the other
hand, the principal directions are themselves spatially-dependent, the
noise-induced drift may be also affected by their rotation. In this
work we derive an expression for the noise-induced drift in two
dimensional anisotropic systems [Eq.~(\ref{eq:higherdrift4})]. The
derivation indeed reveals a new term representing an additional novel
contribution to the noise-induced drift arising from spatial
variations in the directions of the principal axes. The magnitude of
the new term is proportional to the difference between the the
principal diffusion coefficients and the rate of spatial change of the
principal unit vectors (curvature). The newly derived expression for
the noise-induced drift is tested and validated by using Langevin
dynamics simulations of model systems with anisotropic diffusion
tensor. To demonstrate the importance of the drift effect, we use the
derived expression to evaluate the variations in the relative distance
between two transmembrane proteins, and find the noise-induced effect
to be surprisingly large.

The paper is organized as follows. In section \ref{sec:derivation}, we
present our derivation for the noise-induced drift. Section
\ref{sec:simulations}) present results of computer simulations of
Langevin dynamics in two-dimensional anisotropic systems. In section
\ref{sec:proteins} we apply the newly-derived expression to the case
study of pair-diffusion of membrane proteins. Finally, in section
\ref{sec:discussion}, we discuss the results and explain why the
variations of the principal diffusion axes induce an additional
component to the noise-induced drift.

\section{Derivation}
\label{sec:derivation}

\subsection{Noise-induced drift in one dimension}

The displacement, $\Delta x$, of a particle initially located at
$x=x_0$ can be calculated by integrating the full (underdamped)
Langevin equation~(\ref{eq:langevineq}) over the time interval from
$t=0$ to $t=\Delta t$, and taking the average of the different terms
with respect to all possible noise realizations and all possible
values of the initial velocity (or, equivalently, over an ensemble of
particles). This yields the following equation
\begin{eqnarray}
  \left\langle\int_{x_0}^{x_0+\Delta x} \alpha(x)dx\right\rangle=
  -\left\langle m\Delta v\right\rangle
  +\left\langle\int_0^{\Delta t}\beta(t)dt\right\rangle.
  \label{eq:intlangevin}
\end{eqnarray}
Both terms on the r.h.s.~of Eq.~(\ref{eq:intlangevin}) vanish for the
following reasons: The first term is the average momentum change of
the particles, which are moving at constant temperature and experience
no deterministic force due to a potential gradient. Their momentum
distribution function, therefore, remains unchanged and is given by
the symmetric Maxwell-Boltzmann equilibrium distribution.  The second
term represents the average momentum change due to the thermal
noise. It vanishes because the ensemble average at each time instance
and coordinate $\langle\beta(x(t)\rangle=0$ \cite{gjf1,gjf2}. (It is
the noise variance rather than the mean that depends on the coordinate
$x$). We, thus, conclude that
\begin{eqnarray}
  \left\langle\int_{x_0}^{x_0+\Delta x} \alpha(x)dx\right\rangle=0.
  \label{eq:zerofriction}
\end{eqnarray}
Assuming that $\alpha(x)$ is a smooth function which does not change
considerably during the time interval $\Delta t$, one can use the
truncated Taylor expansion $\alpha(x)\simeq
\alpha(x_0)+\alpha^{\prime}(x_0)(x-x_0)$ in (\ref{eq:zerofriction}) to
arrive at the following relation
\begin{eqnarray}
  \langle\Delta x\rangle=-\frac{\alpha^{\prime}(x_0)}{2\alpha(x_0)}
  \left\langle\left(\Delta x\right)^2\right\rangle
  \label{eq:drift1}
\end{eqnarray}
between the mean displacement and the mean-squared displacement (MSD).
Further assuming that the time interval of interest $\Delta t$ is much
larger than the ballistic time $\tau\sim m/\alpha(x_0)$, the MSD on
the r.h.s.~of Eq.~(\ref{eq:drift1}) can be approximated to leading
order by $\left\langle\left(\Delta x\right)^2\right\rangle\simeq
2D(x_0)\Delta t$ which, together with Einstein's relation
(\ref{eq:einstein}), yields
\begin{eqnarray}
  \langle \Delta x\rangle\simeq D^{\prime}(x_0)\Delta t.
  \label{eq:drift2}
\end{eqnarray}
From Eq.~(\ref{eq:drift2}) we identify the drift velocity $v^{\rm
  drift}\equiv \langle \Delta x\rangle/\Delta t$ as being equal to the
gradient of the diffusion coefficient, $D^{\prime}(x)$.

\subsection{multidimensional systems}

In dimensions higher than one, Langevin's equation takes the tensorial
form \cite{lax}
\begin{eqnarray}
  m\frac{dv_i}{dt}=-\alpha\left(\left\{x_k\right\}\right)_{ij}v_j
  +\beta_i\left(x\left( t\right)\right),
  \label{eq:langevinten}
\end{eqnarray}
where the subscripts $i$, $j$, and $k$ denote Cartesian coordinates
and Einstein's summation rule over repeated indices is assumed. The
components of the friction tensor, $\alpha_{ij}$, may depend on all
the space coordinates, $\left\{x_k\right\}$, and the noise satisfies
$\langle\beta_i(t)\rangle=0$, and,
$\langle\beta_i(t)\beta_j(t^{\prime})\rangle=2\alpha_{ij}k_BT\delta(t-t^{\prime})$.
The space-dependent diffusion tensor,
$D_{ij}\left(\left\{x_k\right\}\right)$, is related to $\alpha_{ij}$
via Einstein's relation (\ref{eq:einstein}), which in dimensions
higher than one reads $\alpha_{ik}D_{kj}=k_BT\delta_{ij}$, where
$\delta_{ij}$ is the Kronecker's delta (identity matrix)
\cite{lax}. The drift can be calculated by repeating the derivation
outlined above for one-dimensional systems. This leads to the
generalized form of Eq.~(\ref{eq:drift1})
\begin{eqnarray}
  \label{eq:higherdrift1}
  \alpha_{ij}\langle\Delta x_j\rangle&=&-\frac{\partial \alpha_{ij}}{\partial x_k}
  \left\langle\int\left(x_k-x_k\left(0\right)\right)dx_j\right\rangle\\
  &\simeq& 
  -\frac{\partial \alpha_{ij}}{\partial x_k}
  \left\langle\frac{\Delta x_k\Delta x_j}{2}\right\rangle \simeq
  -\frac{\partial \alpha_{ij}}{\partial x_k}D_{kj}\Delta t. \nonumber
\end{eqnarray}
From Einstein's relation we deduce that $(\partial
\alpha_{ij}/\partial x_k)D_{kj}+\alpha_{ij}(\partial D_{kj}/\partial
x_k)=0$, and by using this result in (\ref{eq:higherdrift1}), we
arrive at
\begin{eqnarray}
 \langle \Delta x_i\rangle \simeq  
\left .\frac{\partial
    D_{ij}\left(\left\{x_k\right\}\right)}{\partial x_j}\right|_0\Delta
  t,
\label{eq:higherdrift2}
\end{eqnarray}
which generalizes Eq.(\ref{eq:drift2}) for $d-$dimensional ($d>1$)
systems.

Since $\alpha_{ij}$ is a real symmetric matrix at each point in space,
it can be diagonalized. Along the {\em local}\/ principal axes, the
friction tensor reads $\alpha_{ij}=\alpha_j\delta_{ij}$ (Einstein's
summation convention is suppressed henceforth), where
$\alpha_j\left(\left\{x_k\right\}\right)=k_BT/D_j\left(\left\{x_k\right\}\right)$
is the coordinate-dependent eigenvalue of the matrix $\alpha_{ij}$
that is associated with the $j$-th principal direction, and $D_j$ is
the corresponding, space-dependent, eigenvalue of the diffusion
matrix. We now consider the Brownian motion of a particle in a
two-dimensional anisotropic system with $D_1(x_1,x_2)\neq
D_2(x_1,x_2)$, where the Cartesian axes of the lab frame, $\hat{x}_1$
and $\hat{x}_2$, are chosen to lie in the {\em initial}\/ principal
directions. The diffusion tensor of the system, $D(x_1,x_2)$, is given
by
\begin{eqnarray}
\left(\begin{array} {lr} D_1\cos^2\theta+D_2\sin^2\theta  & \Delta D\sin 2\theta
  \\ \Delta D\sin 2\theta & D_1\sin^2\theta+D_2\cos^2\theta
\end{array}\right),
\label{eq:difftensor}
\end{eqnarray} 
where $\Delta D\,(x_1,x_2)=D_1-D_2$, and $\theta\,(x_1,x_2)$ is the
angle between the lab frame axes and the local principal directions
[$\theta(t=0)=0$]. We note that the same form
Eq.~(\ref{eq:difftensor}) is also used to study the Brownian motion of
an ellipsoid in a homogeneous medium \cite{han}. However, in the
latter example, the diffusion tensor does not depend on the center of
mass coordinates of the particle as in the present study, but on the
instantaneous orientation of the (non-spherical) Brownian
particle. This implies an important difference between the two
problems. In the case of diffusion of particles with anisotropic
shapes, the translational and orientational degrees of freedom are
decoupled and, therefore, at sufficiently long times the motion must
become isotropic. In the case discussed herein of diffusion in
anisotropic medium, the changes in the orientation of the diffusion
tensor depends on the the position of the particle, which introduces a
strong coupling between the translational diffusion and the
orientational variations.

\begin{figure}[t]
\centering\includegraphics[width=0.4\textwidth]{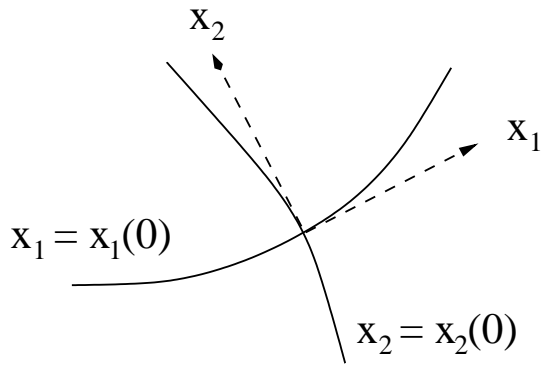} 
\caption{The local principal directions define an orthogonal
  curvilinear coordinate system with unit vectors $\hat{x}_i$
  $(i=1,2)$ that are tangent to the coordinate curves $x_i=x_i(0)={\rm
    Const}$.}
\label{fig:fig1}
\end{figure}

Using Eq.~(\ref{eq:difftensor}) in Eq.~(\ref{eq:higherdrift2}) yields
the following expressions for drift velocity, $v_i^{\rm
  drift}=\langle\Delta x_i\rangle/\Delta t$:
\begin{eqnarray}
v_1^{\rm drift}&=&\frac{\partial D_1}{\partial x_1}+\Delta
D\frac{\partial \theta}{\partial
  x_2} \label{eq:higherdrift3a}\\ v_2^{\rm drift}&=&\frac{\partial
  D_2}{\partial x_2}+\Delta D\frac{\partial \theta}{\partial x_1}.
\label{eq:higherdrift3b}
\end{eqnarray}
The second terms on the right hand sides of
Eqs.~(\ref{eq:higherdrift3a}) and (\ref{eq:higherdrift3b}) can be
reexpressed in a more illuminating form as follows. The unit vectors
$\hat{x}_1$ and $\hat{x}_2$ in the local principal directions define
an orthogonal curvilinear coordinate system, as exemplified in
Fig.~\ref{fig:fig1}. The partial derivatives of $\theta$, appearing in
Eqs.~(\ref{eq:higherdrift3a}) and (\ref{eq:higherdrift3b}), give the
curvatures $c_1$ and $c_2$ of the coordinate curves $x_1=x_1(0)$ and
$x_2=x_2(0)$, respectively: $c_i=\partial \theta/\partial x_i$. In
vector notation (see Fig.~\ref{fig:fig1}): $\vec{c}_1=(\partial
\theta/\partial \hat{x}_1)=-c_1\hat{x}_2$, and $\vec{c}_2=(\partial
\theta/\partial \hat{x}_2)=c_2\hat{x}_1$, which allows writing
Eq.~(\ref{eq:higherdrift2}) in the following vectorial form:
\begin{eqnarray}
  \left(\begin{array} {l} v_1^{\rm drift} \\ v_2^{\rm drift}\end{array}\right)=
  \left(\begin{array} {l}
    \partial D_1/\partial x_1 \\
    \partial D_2/\partial x_2
  \end{array}\right)+
  \left(\begin{array} {l} (D_1-D_2)c_2
    \\ (D_2-D_1)c_1\end{array}\right),
    \label{eq:higherdrift4}
\end{eqnarray}   
which constitutes the main new result of this paper.

\section{Langevin dynamics simulations}
\label{sec:simulations}

\begin{figure}[t]
\centering\includegraphics[width=0.45\textwidth]{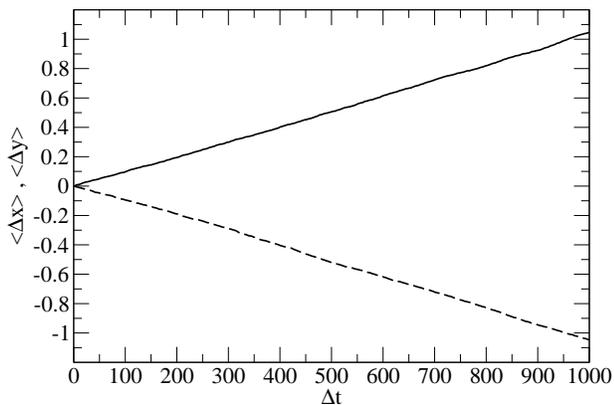} 
\caption{The mean displacements $\langle\Delta x_1\rangle$ (solid
  line) and $\langle\Delta x_2\rangle$ (dashed line) as a function of
  time, computed from Langevin dynamics simulations of $5\times 10^5$
  stochastic trajectories of a particle moving in an anisotropic
  system with $D_1=10x_1/(x_2)^2$ and $D_2=10/x_2$, and starting at
  $x_1=x_2=100$. }
\label{fig:fig2}
\end{figure}

The two vectors on the r.h.s.~of Eq.~(\ref{eq:higherdrift4}) depict
two distinct drifting effects. The former represents the
$d$-dimensional generalization of the one-dimensional equation
(\ref{eq:drift2}) for the drift in the direction of increasing
diffusivity.  Its validity can be tested by considering an example
where the principal directions of the diffusion tensor are fixed, in
which case the angle $\theta$ is constant and the second vector is
null. This is illustrated in Fig.~\ref{fig:fig2}, which shows results
for $\langle\Delta x_1\rangle$ and $\langle\Delta x_2\rangle$ vs.~time
in an anisotropic two-dimensional system with $D_1=10x_1/(x_2)^2$ and
$D_2=10/x_2$. The results are based on inertial Langevin dynamics
simulations of $5\times 10^5$ trajectories of a particle of unity mass
($m=1$) at constant temperature ($k_BT=1$) starting at
$(x_1,x_2)=(100,100)$. The trajectories are computed with $dt=10^{-4}$
(which is three orders of magnitude smaller than the ballistic time),
using a method based on the robust G-JF Langevin thermostat
\cite{gjf3,gjf4} and the novel ``inertial'' convention for assigning
the values of the friction function at each time step \cite{gjf1}:
$\alpha_i^{\rm
  inertial}\equiv\left[\alpha_i\left(\left\{x_k(t)\right\}\right)+
  \alpha_i\left(\left\{x_k(t)+v_k(t)dt\right\}\right)\right]/2$.  From
Eq.~(\ref{eq:higherdrift4}), we expect the drift velocities in this
case to be: $v_1^{\rm drift}\simeq\partial_{x_1}D_1(100,100)=10^{-3}$
and $v_2^{\rm drift}\simeq\partial_{x_2}D_2(100,100)= -10^{-3}$, which
is in agreement with the data in Fig.~\ref{fig:fig2}.

\begin{figure}[t]
\centering\includegraphics[width=0.45\textwidth]{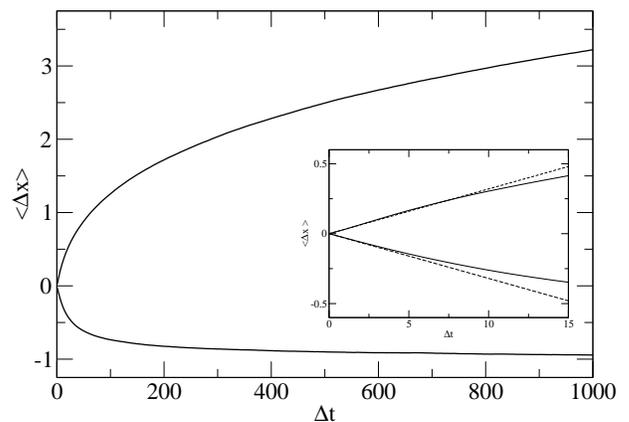} 
\caption{The mean displacement in the $x$ direction, $\langle\Delta
  x\rangle$, as a function of time. The monotonically increasing
  (decreasing) curve depicts results for a particle starting at
  $(x,y)=(1,0)$ and moving in an anisotropic system with $D_{r}=1/25$,
  and $D_{\theta}=1/125$ ($D_{r}=1/125$, and $D_{\theta}=1/25$). The
  inset shows a magnification of the initial region $\Delta t\leq 15$,
  where the tangent dashed lines depict the asymptotic behavior
  $\langle \Delta x \rangle \sim \pm (4/125) \Delta t$, expected from
  Eq.~(\ref{eq:higherdrift4}).}
\label{fig:fig3}
\end{figure}

The second vector on the r.h.s.~of Eq.~(\ref{eq:higherdrift4}) depicts
a different drifting effect, namely the one which is associated with
the spatial variations in the principal directions of the diffusion
tensor. In order to focus on this contribution to the mean
displacement of the particle, we consider a system with constant
diffusion coefficients $D_1$ and $D_2$ ($D_1\neq D_2$), in which case
the first vector is null. As examples, we consider two
radially-symmetric systems - one with diffusion coefficients
$D_1=D_r=1/25$ and $D_2=D_{\theta}=1/125$, in the radial and
tangential directions, respectively, and the other with
$D_1=D_r=1/125$ and $D_2=D_{\theta}=1/25$. Using a method similar to
that employed in Fig.~\ref{fig:fig2} (see footnote~\cite{footnote1}),
we compute $5\times 10^5$ trajectories, each of which starting at
$(x,y)=(1,0)$. Fig.~{\ref{fig:fig3}} depicts the mean displacement
along the $x$ axis, which is the initial radial direction, as a
function of time. The monotonically increasing curve shows the mean
displacement of the particle in the system where $D_r=1/25$ and
$D_{\theta}=1/125$, whereas the decreasing curve corresponds to the
system with $D_r=1/125$ and $D_{\theta}=1/25$. In the former case, the
particle drifts outward ($\langle \Delta x\rangle >0$), which is in
agreement with Eq.~(\ref{eq:higherdrift4}) for
$D_r=D_1>D_2=D_{\theta}$. Conversely, the particle moves inward for
$D_r=D_1<D_2=D_{\theta}$, which is indeed observed in the
simulations. Notice that in the latter case, the displacement
$\langle\Delta x\rangle\rightarrow -1$ for $\Delta t\rightarrow
\infty$, which is anticipated since this is exactly the initial
distance of the particle from the symmetry center of the system. At
small times (i.e., when the particle is still close to the point of
origin), we expect the drift velocity $v_1^{\rm drift}$ to converge to
the asymptotic values of $(D_1-D_2)c_2=\pm(4/125)$ [see
  Eq.~(\ref{eq:higherdrift4})]. This result is captured by the
simulations, as demonstrated in inset to Fig.~{\ref{fig:fig3}} by the
tangent dashed lines depicting the asymptotic behavior $\langle \Delta
x \rangle \sim \pm (4/125) \Delta t$. For both systems (data not
shown), the mean displacement along the initial angular direction
$\langle\Delta y\rangle=0$, which is expected from symmetry arguments,
and is also consistent with Eq.~(\ref{eq:higherdrift4}) considering
that the curvature $c_1$ of $r$-lines is zero.

\section{Pair diffusion of transmembrane proteins}
\label{sec:proteins}

A particularly interesting example of two-dimensional diffusion with
radial symmetry is the relative pair-diffusion between two
transmembrane proteins. This, and other closely-related setups, have
attracted a considerable attention because the lateral diffusion of
membrane proteins and lipid domains is an important biophysical factor
in controlling the dynamics and functioning of the cell membrane (see
reviews \cite{brown,andelman}, and references therein). Let us
consider two membrane inclusions with cylindrical cross-sections of
radius $a$, and denote by $\vec{r}$ the vector distance between them
($r=|\vec{r}|\gg a$). The temporal evolution of $\vec{r}=(r,\theta)$
is diffusive, and the associated diffusion coefficients vary with $r$
because of the hydrodynamic interactions between the
proteins. Explicitly, the radial and tangential diffusion coefficients
of the vector $\vec{r}$ are given by \cite{diamant}
\begin{eqnarray}
  D_r(r)&=&\frac{k_BT}{2\pi\eta_m}\left[\ln\left(r/a\right)-3/2+\kappa
    r/3\right] \label{eq:sd1}\\
  D_{\theta}(r)&=&\frac{k_BT}{2\pi\eta_m}\left[\ln\left(r/a\right)-1/2-
    \kappa r/3\right], \label{eq:sd2}
\end{eqnarray}
where $\kappa^{-1}$ is the Saffman-Delbr\"{u}ck length given by the
ratio between the two-dimensional membrane viscosity $\eta_m$ and
twice the three-dimensional viscosity of the embedding fluid $\eta_f$:
$\kappa^{-1}=\eta_m/2\eta_f$. For lipid bilayers $\eta_m\sim
10^{-10}-10^{-9}\ {\rm Pa\cdot sec\cdot m}$, implying that
$\kappa^{-1}\sim 0.1-1\ \mu{\rm m}$.  Using expressions (\ref{eq:sd1})
and (\ref{eq:sd2}) in Eq.~(\ref{eq:higherdrift4}) gives the average
velocity at which the inclusions are moving away of each other:
$v_r^{\rm drift}=k_BT\kappa/(2\pi\eta_m)\sim 1-10^2\ \mu{\rm m/sec}$,
which is (i) independent of $r$, and (ii) surprisingly large
considering that the linear size of cellular membranes is typically
$\sim 10\ \mu{\rm m}$. There are obviously many other factors that
produce an inhomogeneous diffusion environment for a membrane protein,
most notably the presence of viscous raft domains and the cytoskeleton
meshwork that generates corrals in which the protein may be localized
\cite{nicolson,kusumi}. Nevertheless, the magnitude of the
noise-induced drift (as indicated by the above calculation of
$v_r^{\rm drift}$) appears to be quite large, which suggests that the
hydrodynamic interactions between the proteins cannot be neglected
when studying their lateral diffusion in an inhomogeneous membrane. We
remind here (see earlier discussion) that in isothermal systems, the
drifting effect is countered an opposite trapping effect in regions of
low diffusivity. Therefore, when two proteins come within close
vicinity to each other, they would tend to remain closely separated
\cite{metzler}, and this tendency is likely to be intensified by
shorter range attractive interactions, for instance van der Waals and
membrane-mediated interactions \cite{deserno}.

We note that expressions (\ref{eq:sd1}) and (\ref{eq:sd2}) hold in the
regime where $r\ll \kappa^{-1}$. For $r\gg\kappa^{-1}$, the radial and
azimuthal diffusion coefficients are given by \cite{diamant}
  \begin{eqnarray}
  D_r(r)&=&\frac{k_BT}{2\pi\eta_m}\left[\ln\left(2/\kappa
    a\right)-\gamma-2/\left(\kappa
    r\right)\right] \label{eq:sd3}\\
  D_{\theta}(r)&=& \frac{k_BT}{2\pi\eta_m}\left[\ln\left(2/\kappa
    a\right)-\gamma-2/\left(\kappa
    r\right)^2\right], \label{eq:sd4}
\end{eqnarray}
where $\gamma\simeq 0.58$ is Euler's constant. Using these expressions
in Eq.~(\ref{eq:higherdrift4}), we arrive at $v_r^{\rm
  drift}=k_BT/(\pi\eta_m\kappa^2 r^3)$, for the noise-induced radial
drift velocity at large separations.

\section{Discussion}
\label{sec:discussion}

In this paper we studied the problem of single particle diffusion in
multidimensional systems with space-dependent diffusion coefficient,
focusing on the noise-induced drift in two dimensions. By following
the derivation for the drift in one-dimensional systems
[Eq.~(\ref{eq:drift2})], we arrive at the generalized form
(\ref{eq:higherdrift2}) in higher dimensions. This expression is
further analyzed in two-dimensions by considering a coordinate system
that aligns along the local principal diffusion axes. This analysis
yields Eq.~(\ref{eq:higherdrift4}), where the noise-induced drift
vector is expressed as the sum of two terms representing two distinct
noise-induced drifting effects. The first effect, which arises also in
the one-dimensional case, originates from the fact that the ballistic
distance grows proportionally to the local diffusion coefficient and,
therefore, the particle moves more persistently (``makes larger
steps'') in the direction of increasing diffusivity. The second effect
is a novel one, existing only in dimensions larger than one. It stems
from the gradual rotation of the principal diffusion directions,
$\hat{x}_1$ and $\hat{x}_2$ occurring during the motion of the
particle, and can be understood as follows. At very large times, after
the particle moves to distant regions and the memory of the initial
principal directions is lost, the dynamics becomes isotropic, and is
characterized by the average diffusion coefficient
$(D_1+D_2)/2$. Assuming (without loss of generality) that $D_1>D_2$,
this implies that as the particle diffuses away from its initial
position, the effective diffusion coefficient in the initial
$\hat{x}_1$ direction decreases, $D_1^{\rm eff}<D_1$, while the
diffusion along the initial $\hat{x}_2$ direction occurs with an
effectively increasing diffusion coefficient, $D_2^{\rm
  eff}>D_2$. {\em The origin of the drift lies in the fact that the
  rate of rotation is spatially-dependent.}\/ It is larger in the
direction where the curvature of the $x_1$ and $x_2$ curves increases,
which explains the form of the second vector in expression
(\ref{eq:higherdrift4}) for the two-dimensional noise-induced drift
velocity.

We used the newly-derived expression to estimate the drift velocity of
a pair of membrane proteins, and found it to be surprisingly
large. The actual diffusive dynamics of membrane proteins is obviously
more complex. It takes place in a highly-crowded environment which
implies that many-body effects are important. Nevertheless, our study
clearly highlights the fact that hydrodynamic interactions between
proteins (which are the origin of the diffusivity spatial variations
here) are likely to be key factors.  \\

{\bf Acknowledgments:} I thank Haim Diamant for helpful discussions on
diffusion of membrane inclusions and for critical reading of the
manuscript. This work was supported by the Israel Science Foundation
(ISF) through grant number 1087/13.


\begin{thebibliography}{99}

\bibitem{volpe1} T. Brettschneider, G. Volpe, L. Helden, J. Wehr, and
  C. Bechinger, Phys. Rev. E {\bf 83}, 041113 (2011). 

\bibitem{brenner1} H. Brenner, Chem. Eng. Sci. {\bf 16}, 242, (1961).

\bibitem{brenner2} A.J. Goldman, R.G. Cox, and H. Brenner,
  Chem. Eng. Sci. {\bf 22}, 637 (1967).

\bibitem{tauber} B. Schulz, D. T\"{a}uber, J. Schuster,
  T. Baumg\"{a}rtel, and C. von Borczyskowski, Soft Matter {\bf 7},
  7431 (2011).
  
\bibitem{pumpa} M. Pumpa and F. Cichos, J. Phys. Chem. B {\bf 116},
  14487 (2012).

\bibitem{perrin} F.  Perrin, J. Phys. Radium {\bf 5}, 497 (1934).
  
\bibitem{han} Y. Han, A. M. Alsayed, M. Nobili, J. Zhang,
  T. C. Lubensky, and A. G. Yodh, Science {\bf 314}, 626 (2006).

\bibitem{volpe2} G. Volpe and J. Wehr, Rep. Prog. Phys. {\bf 79},
    053901 (2016).

\bibitem{langevin} P. Langevin, C. R. Acad. Sci. (Paris) {\bf 146},
  530 (1908).

\bibitem{kubo} R. Kubo, Rep. Prog. Phys. {\bf 29}, 255 (1966).

\bibitem{gjf1} O. Farago and N. Gr\o nbech-Jensen, J. Stat. Phys. {\bf
156}, 1093 (2014).

\bibitem{mannella} R. Mannella and V. P. E. McClintock, Fluct. Noise
  Lett. {\bf 11}, 1240010 (2012).

\bibitem{lau} A. W. C. Lau and T. C. Lubensky, Phys. Rev. E {\bf 76},
011123 (2007).

\bibitem{gjf2} O. Farago and N. Gr\o nbech-Jensen, Phys. Rev. E {\bf
89}, 013301 (2014).
 
\bibitem{lancon} P. Lan\c{c}on, G. Batrouni, L. Lobry, and N. Ostrowsky,
  Europhys. Lett. {\bf 54}, 28 (2001).

\bibitem{lax} M. Lax, Rev. Mod. Phys. {\bf 32}, 25 (1960).

\bibitem{gjf3} N. Gr\o nbech-Jensen, and O. Farago, Mol. Phys. {\bf
   111}, 983 (2013).

\bibitem{gjf4} N. Gr\o nbech-Jensen, N. R. Hayre, and O. Farago,
Comput. Phys. Commun. {\bf 185}, 524 (2014).

\bibitem{footnote1} The simulations use the G-JF algorithm of
  {\protect ref.~\cite{gjf3}} with $dt=10^{-4}$, where at each time
  step the coordinates and velocities are updated along the
  instantaneous principal directions. Denoting by $\theta$ the angle
  between the radial direction and the $x$-axis, the angular
  variations are accounted for by using an ``inertial'' convention for
  the angle, i.e., by considering for each time step the angle
  $\theta^{\rm
    inertial}\equiv\left[\theta\left(\left\{x_k(t)\right\}\right)+
    \theta\left(\left\{x_k(t)+v_k(t)dt\right\}\right)\right]/2$.

\bibitem{brown} F. L. H. Brown, Quart. Rev. Biophys. {\bf 44}, 391
  (2011).

\bibitem{andelman} S. Komura and D. Andelman, Adv. Colloid Interface
  Sci. {\bf 208}, 34 (2014).

\bibitem{diamant} N. Oppenheimer and H. Diamant, Biophys. J. {\bf 96},
  3041 (2009).

\bibitem{nicolson} G. L. Nicolson, Biochim. Biophys. Acta {\bf 1838},
  1451 (2014).
  
\bibitem{kusumi} A. Kusumi, K. G. N. Suzuki1, R. S. Kasai, K. Ritchie,
  and T. K. Fujiwara, Trends Biochem. Sci. {\bf 36}, 604 (2011).

\bibitem{metzler} J-H Jeon, M. Javanainen, H. Martinez-Seara,
  R. Metzler, and I. Vattulainen, Phys. Rev. X {\bf 6}, 021006 (2016).
  
\bibitem{deserno} M. Deserno, K. Kremer, H. Paulsen, C. Peter, and
  F. Schmid, in {\it Adv. Polym. Sci. - From Single Molecules to
    Nanoscopically Structured Materials}\/, Vol. 260 (Eds. T. Basche,
  K. M?llen, and M. Schmidt), pp. 237-283 (Springer-Verlag, Berlin,
  2014).
  
\end{thebibliography}
\end{document}